\documentclass[12pt,a4paper]{article}
\usepackage{amsfonts,latexsym}
\usepackage{graphicx,color}
\usepackage{dcolumn}
\usepackage{graphicx}
\usepackage{amsmath}
\usepackage{amsfonts}
\usepackage{amssymb}

\renewcommand{\thefootnote}{\fnsymbol{footnote}}

\begin{document}

\vspace{12mm}

\begin{center}
{{{\Large {\bf Double instability of Schwarzschild black holes\\ in Einstein-Weyl-scalar theory  }}}}\\[10mm]

{Yun Soo Myung$^a$\footnote{e-mail address: ysmyung@inje.ac.kr}}\\[8mm]

{${}^a$Institute of Basic Sciences and Department  of Computer Simulation, Inje University, \\ Gimhae 50834, Korea\\[0pt] }

\end{center}
\vspace{2mm}

\begin{abstract}
We  study the stability of Schwarzschild black
hole in  Einstein-Weyl-scalar (EWS) theory with a quadratic scalar coupling to the Weyl term.
Its linearized theory  admits the Lichnerowicz equation for Ricci tensor as well as  scalar equation.
The linearized  Ricci-tensor carries with a regular mass term ($m^2_2$), whereas the linearized scalar has a tachyonic  mass term ($-1/m^2_2$).
It turns out that the double instability  of Schwarzschild black hole in EWS theory  is given by  Gregory-Laflamme  and  tachyonic instabilities.
In the small mass  regime of $m_2<0.876$, the Schwarzschild black hole becomes unstable against Ricci-tensor perturbations,
while tachyonic instability is achieved for  $m_2<1.174$. The former would provide a single branch of scalarized black holes, whereas the latter would induce  infinite  branches of scalarized black holes.

\end{abstract}
\vspace{5mm}

\vspace{1.5cm}

\hspace{11.5cm}
\newpage
\renewcommand{\thefootnote}{\arabic{footnote}}
\setcounter{footnote}{0}


\section{Introduction}
Recently, black hole solutions with scalar hair obtained  from Einstein-Gauss-Bonnet-scalar (EGBS) theories~\cite{Antoniou:2017acq,Doneva:2017bvd,Silva:2017uqg} and Einstein-Maxwell-scalar theory~\cite{Herdeiro:2018wub} have received much attention
 because they have  uncovered  easily an evasion of  the no-hair theorem~\cite{Bekenstein:1995un} by introducing a non-minimal (quadratic) scalar coupling function $f(\phi)$ to Gauss-Bonnet and Maxwell terms.
We note  that these scalarized black hole solutions  are closely related to the appearance of  tachyonic instability for bald black holes.
In these linearized theories, the instability of Schwarzschild black hole is determined solely by the linearized scalar equation where the Gauss-Bonnet term acts as an effective mass term~\cite{Myung:2018iyq}, while
the instability of Reissner-Nordstr\"{o}m (RN) black hole is given just  by the linearized scalar equation where the Maxwell term plays the role of  an effective mass term~\cite{Myung:2018vug}.
This is allowed because their linearized Einstein and Einstein-Maxwell equations reduce to those for the  linearized Einstein theory around Schwarzschild black hole and the Einstein-Maxwell theory around RN black hole, which turned out to be stable against tensor (metric) and vector-tensor perturbations.

It was well known that a higher curvature gravity (Einstein-Weyl theory) with a mass coupling parameter $m^2_2$  has  provided  the non-Schwarzschild black hole solution which crosses the Schwarzschild black hole solution at the bifurcation point of $m_2=0.876$~\cite{Lu:2015cqa}.
This solution indicates  the black hole with non-zero Ricci tensor ($\bar{R}_{\mu\nu}\not=0$), comparing to zero Ricci tensor ($\bar{R}_{\mu\nu}=0$) for Schwarzschild black hole.
We note that the trace no-hair theorem for Ricci scalar played an important role in obtaining the non-Schwarzschild black hole solution.
It is worth noting that the instability of Schwarzschild black hole was found in the massive gravity theory~\cite{Babichev:2013una,Brito:2013wya} since the Schwarzschild black hole was known to be dynamically stable against tensor perturbations in Einstein theory~\cite{Regge:1957td,Zerilli:1970se}.
In the linearized Einstein-Weyl theory, the instability bound of Schwarzschild black hole was  found as $m_2<0.876$ with $r_+=1$  when solving  the Lichnerowicz equation for the linearized Ricci tensor~\cite{Myung:2013doa}, which is the same equation as the linearized Einstein equation around a (4+1)-dimensional black string where the Gregory-Laflamme (GL) instability appeared firstly~\cite{Gregory:1993vy}.
A little  difference is that the instability of Schwarzschild black hole  arose from the massiveness of $m_2\not=0$ in the Einstein-Weyl theory, whereas the GL instability appeared from the geometry of an extra $z$ dimension in (4+1)-dimensional black string theory. This means that the mass $m_2$ trades for the extra dimension $z$.

In the present work, we wish to study two instabilities of Schwarzschild  black holes simultaneously  by introducing   the Einstein-Weyl-scalar theory with a quadratic scalar coupling  to  Weyl term, instead of Gauss-Bonnet term. In this case,  the linearized  Ricci-tensor $\delta R_{\mu\nu}$ has a regular mass term $m^2_2$, whereas the linearized scalar $\delta \phi$ possesses a tachyonic  mass term ($-\frac{1}{m^2_2}$).
The linearized scalar equation around Schwarzschild black hole undergoes tachyonic instability for $m_2<1.174$, while  the Lichnerowicz equation for linearized  Ricci-tensor reveals  GL instability for $m_2<0.876$.
We expect that the former may induce infinite branches ($n=0,1,2,\cdots$) of scalarized black holes, while the latter admits a single branch ($m_2\not=0$) of scalarized black holes.
This means that their role of the  mass term are quite different for producing scalarized black holes.

\section{Einstein-Weyl-scalar (EWS) theory} \label{sec1}

We introduce the EWS theory defined by
\begin{equation}
S_{\rm EWS}=\frac{1}{16 \pi}\int d^4 x\sqrt{-g}\Big[ R-2\partial_\mu \phi \partial^\mu \phi-\frac{f(\phi)}{2m^2_2} C^2\Big],\label{Action}
\end{equation}
where $f(\phi)=1+\phi^2$ is  a quadratic scalar coupling function, $m_2^2$ denotes a mass coupling parameter, and $C^2$ represents the Weyl term (Weyl scalar invariant) given  by
\begin{equation}
C^2(\equiv C_{\mu\nu\rho\sigma}C^{\mu\nu\rho\sigma})=2(R_{\mu\nu}R^{\mu\nu}-\frac{R^2}{3})+{\cal R}_{\rm GB}^2\label{Action2}
\end{equation}
with the Gauss-Bonnet term  ${\cal R}_{\rm GB}^2=R^2-4R_{\mu\nu}R^{\mu\nu}+R_{\mu\nu\rho\sigma}R^{\mu\nu\rho\sigma}$. In the limit of $m_2^2\to \infty$,  the Weyl term decouples and the theory reduces to the tensor-scalar theory.
We wish to emphasize that  scalar couplings to Gauss-Bonnet term were mostly used to find black holes with scalar hair within EGBS theory because it provides an effective mass term for a linearized scalar without modifying  metric perturbations~\cite{Antoniou:2017acq,Doneva:2017bvd,Silva:2017uqg}. This is so because the Gauss-Bonnet term is a topological term in four dimensions.
Actually, the Weyl term is similar to the Maxwell term ($F^2$) because both they  are conformally invariant and their variations with respect to $g_{\mu\nu}$  are traceless.

From the action (\ref{Action}), we derive  the Einstein  equation
\begin{eqnarray}
 G_{\mu\nu}=2\partial _\mu \phi\partial _\nu \phi -(\partial \phi)^2g_{\mu\nu}+\frac{2(1+\phi^2)B_{\mu\nu}}{m^2_2}-\frac{\Gamma_{\mu\nu}}{m^2_2}, \label{equa1}
\end{eqnarray}
where $G_{\mu\nu}=R_{\mu\nu}-(R/2)g_{\mu\nu}$ is  the Einstein tensor.
Here, $B_{\mu\nu} (B^\mu~_\mu=0)$ coming  from the first part of (\ref{Action2}) is the Bach tensor  defined as
\begin{eqnarray}
B_{\mu\nu}&=& R_{\mu\rho\nu\sigma}R^{\rho\sigma}-\frac{g_{\mu\nu}}{4} R_{\rho\sigma}R^{\rho\sigma}- \frac{R}{3}\Big(R_{\mu\nu}-\frac{g_{\mu\nu}}{4}R\Big) \nonumber \\
&+& \frac{1}{2}\Big(\nabla^2R_{\mu\nu}-\frac{g_{\mu\nu}}{6}\nabla^2 R-\frac{1}{3} \nabla_\mu\nabla_\nu R\Big) \label{bach}
\end{eqnarray}
and  $\Gamma_{\mu\nu}$   is given by
\begin{eqnarray}
\Gamma_{\mu\nu}&=&-\frac{4}{3}R\nabla_{(\mu} \Psi_{\nu)}-\nabla^\alpha \Psi_\alpha \Big(3R_{\mu\nu}-\frac{4g_{\mu\nu}}{3}R\Big)+ 6R_{(\mu|\alpha|}\nabla^\alpha \Psi_{\nu)} \nonumber \\
&-&3 R^{\alpha\beta}\nabla_\alpha\Psi_\beta g_{\mu\nu}
+4R^{\beta}_{~\mu\alpha\nu}\nabla^\alpha\Psi_\beta  \label{equa2}
\end{eqnarray}
with
\begin{equation}
\Psi_{\mu}= 2\phi \partial_\mu \phi.
\end{equation}
Its trace is not zero as  $\Gamma^\mu~_\mu=R\nabla^\rho\Psi_\rho-2R^{\rho\sigma}\nabla_\rho\Psi_\sigma$.

Importantly, the scalar  equation is given by
\begin{equation}
\nabla^2 \phi +\frac{C^2}{4m^2_2} \phi=0 \label{s-equa}.
\end{equation}

Considering  $\bar{\phi}=0$,  the Schwarzschild  solution is found  from Eqs.(\ref{equa1}) and (\ref{s-equa}) as
\begin{equation} \label{ansatz}
ds^2_{\rm SBH}= \bar{g}_{\mu\nu}dx^\mu dx^\nu=-\Big(1-\frac{r_+}{r}\Big)dt^2+\frac{dr^2}{\Big(1-\frac{r_+}{r}\Big)}+r^2d\Omega^2_2
\end{equation}
with horizon radius $r_+=2M$. This Schwarzschild background gives us $\bar{R}_{\mu\nu\rho\sigma}\not=0,~\bar{R}_{\mu\nu}=0,$ and $\bar{R}=0$.
In this case, one finds easily that  $\bar{C}^2=\bar{R}_{\mu\nu\rho\sigma}\bar{R}^{\mu\nu\rho\sigma}=\frac{12r_+^2}{r^6}=\bar{\cal R}^2_{\rm GB}$.

\section{Double instability for Schwarzschild black hole}

For the stability analysis of Schwarzschild black hole, we need the two linearized equations which describe the metric perturbation   $h_{\mu\nu}$ in ($g_{\mu\nu}=\bar{g}_{\mu\nu}+h_{\mu\nu}$) and scalar perturbation $\delta \phi$ in ($\phi=0+\delta\phi)$ propagating around (\ref{ansatz}).   They are obtained  by linearizing Eqs.(\ref{equa1}) and (\ref{s-equa}) as
\begin{eqnarray}
 && \bar{\nabla}^2\delta G_{\mu\nu}+2\bar{R}_{\mu\rho\nu\sigma}\delta G^{\rho\sigma}-\frac{1}{3}\Big(\bar{\nabla}_\mu \bar{\nabla}_\nu-\bar{g}_{\mu\nu}\bar{\nabla}^2\Big)\delta R-m^2_2 \delta
  G_{\mu\nu}=0 , \label {lin-eq1}\\
 && \left(\bar{\nabla}^2+ \frac{3r_+^2}{m^2_2r^6}\right)\delta \phi= 0 \label{lin-eq2}
\end{eqnarray}
with $\delta G_{\mu\nu}=\delta R_{\mu\nu}-\delta R \bar{g}_{\mu\nu}/2$ the linearized Einstein tensor. Here, we note that `$m^2_2$' in  Eq.(\ref{lin-eq1}) is regarded as  a regular mass term, while `$3r_+^2/m^2_2r^6$' in Eq.(\ref{lin-eq2}) corresponds to a tachyonic mass term for $m^2_2>0$.
Taking the trace over Eq.(\ref{lin-eq1}) leads to
\begin{equation}
m^2_2 \delta R=0,
\end{equation}
which implies the non-propagation of a linearized  Ricci scalar as
\begin{equation}
\delta R=0. \label{non-RS}
\end{equation}
We confirm Eq.(\ref{non-RS})  by linearizing $R=2(\partial \phi)^2+\Gamma^\mu~_\mu/m^2_2$.
This non-propagation of linearized scalar plays an important role in obtaining a linearized theory of the EWS theory.
Plugging Eq.(\ref{non-RS}) into Eq.(\ref{lin-eq1}), one finds
the Lichnerowicz-Ricci tensor equation for the traceless and transverse Ricci tensor $\delta R_{\mu\nu}$  as
\begin{equation}\label{EOM9}
\Big(\bar{\Delta}_{\rm L}+m^2_2 \Big) \delta R_{\mu\nu}=0,
\end{equation}
where the Lichnerowicz operator on the Schwarzschild background is given by
\begin{equation} \label{lichnero}
\bar{\Delta}_{\rm L} \delta R_{\mu\nu}=-\bar{\nabla}^2\delta R_{\mu\nu}-2\bar{R}_{\mu\rho\nu\sigma}\delta R^{\rho\sigma}.
\end{equation}
Here, we consider  $m^2_2>0$ for non-tachyonic case.
Actually, Eq.(\ref{EOM9}) describes a massive spin-2 mode ($\delta R_{\mu\nu}$) with mass $m_2$  propagating on the  Schwarzschild black hole background.
Let us solve the Lichnerowicz-Ricci tensor equation (\ref{EOM9}) by adopting  $\delta R_{\mu\nu}(t,{\bf x})=e^{\Omega t}\delta \tilde{R}_{\mu\nu}({\bf x})$.
Its $s(l=0)$-mode in polar sector satisfies the Schr\"{o}dinger-type equation when introducing a tortoise coordinate $r_*=\int[dr/(1-r_+/r)]$
\begin{equation} \label{Z-eq}
\frac{d^2\delta\tilde{R}^{l=0}_{\mu\nu}}{dr^2_*}-[\Omega^2+V_{\rm Z}(r)]\delta\tilde{R}^{l=0}_{\mu\nu}=0,
\end{equation}
where the Zerilli potential $V_{\rm Z}(r)$ is given by~\cite{Brito:2013wya,Lu:2017kzi} \label{Z-p}
\begin{equation}
V_{\rm Z}(r)=\Big(1-\frac{r_+}{r}\Big)\Big[m^2_2 +\frac{r_+}{r^3}-\frac{12m^2_2r_+(r-0.5r_+)+6m^4_2r^3(2r_+-r)}{(r_++m^2_2r^3)^2}\Big].
\end{equation}
As is shown in (Left) Fig. 1, all potentials with $m_2\not=0$ induce  negative region near the horizon, while their asymptotic forms are given by $m^2_2>0$.
The negative region becomes wide and deep as the mass parameter $m_2$ decreases, implying GL instability of the Schwarzschild black hole.
In case of $m_2=0$, however,  there is no GL instability because its potential $V_{\rm Z}(r)$ is positive definite outside the horizon.
Solving Eq.(\ref{Z-eq}) numerically  with appropriate boundary conditions, one finds the GL instability bound from (Left) Fig. 2 as
\begin{equation}
0<m_2<m_2^{\rm th}=0.876,\quad  {\rm for}~r_+=1,
\end{equation}
where $m_2^{\rm th}$ denotes threshold of GL instability. It is important to note that this bound is found in the EWS theory, but there is no such bound in the EGBS theory.

\begin{figure*}[t!]
   \centering
   \includegraphics{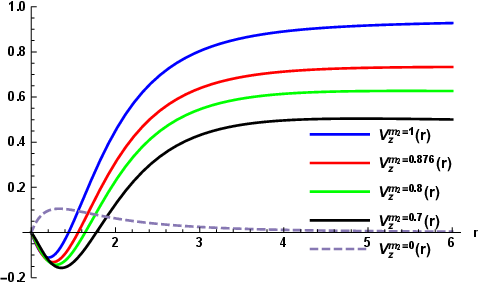}
      \hfill%
    \includegraphics{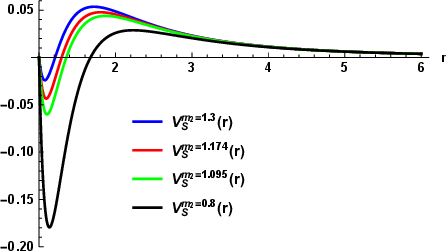}
\caption{(Left) Zerilli potentials $V_{\rm Z}(r)$ as function of $r\in [r_+=1,6]$. $V_{\rm Z}(r)$ is depicted  with  five mass parameters $m_2=1$ (stable), 0.876 (threshold of GL instability),  0.8 (unstable),  0.7  (unstable), and 0 (stable), respectively.
 These all except $m_2=0$ develop negative regions near the horizon. (Right) Scalar potentials $V_{\rm S}(r)$ with four mass parameters $m_2$=1.3 (stable), 1.174 (threshold of tachyonic instability), 1.095 (sufficient condition for tachyonic instability), and  0.8 (unstable). }
\end{figure*}
In the study of   the instability for the  Euclidean Schwarzschild black hole together with  Einstein gravity, Gross, Perry, and Yaffe have found that there is just one normalizable negative-eigenvalue mode of the Licherowicz
operator [$(\Delta^{\rm E}_{\rm L}-\lambda_{\rm GPY})h_{\mu\nu}=0$]~\cite{Gross:1982cv}. This connection could be realized  from Eq.(\ref{EOM9}) because when one considers $\delta R_{\mu\nu}=\bar{\Delta}_{\rm L}h_{\mu\nu}/2$
for $\bar{\nabla}^\mu h_{\mu\nu}=0$ and $h^\mu~_\mu=0$,  Eq.(\ref{EOM9}) implies  that $\bar{\Delta}_{\rm L}h_{\mu\nu}=0$ or $(\bar{\Delta}_{\rm L}+m^2_2)h_{\mu\nu}=0$.
 Its eingenvalue is given by $\lambda_{\rm GPY}[=-(m_2^{\rm th})^2]=-0.768/r_+^2$ which  was noted in the early study of Schwarzschild black hole within  higher curvature gravity~\cite{Whitt:1985ki}. Indeed,  $\lambda_{\rm GPY}$ is related to  the thermodynamic instability of negative heat capacity $C=-2\pi r_+^2$ for Schwarzschild black hole in canonical ensemble.

On the other hand, we focus on the linearized scalar equation (\ref{lin-eq2}) which is the same form as found  in the linearized  EGBS theory.
Considering
\begin{equation} \label{scalar-sp}
\delta \phi(t,r,\theta,\varphi)=\frac{u(r)}{r}e^{-i\omega t}Y_{lm}(\theta,\varphi),
\end{equation}
 the radial equation for $s(l=0)$-mode scalar  leads to the Schr\"{o}dinger-type equation
\begin{equation} \label{scalar-eq}
\frac{d^2u}{dr_*^2}+\Big[\omega^2-V_{\rm S}(r)\Big]u(r)=0,
\end{equation}
where the scalar potential $V_{\rm S}(r)$ is given by
\begin{equation} \label{pot-c}
V_{\rm S}(r)=\Big(1-\frac{r_+}{r}\Big)\Big[\frac{r_+}{r^3}-\frac{3r_+^2}{m^2_2r^6}\Big],
\end{equation}
where the last term corresponds to a tachyonic mass term.
\begin{figure*}[t!]
   \centering
  \includegraphics{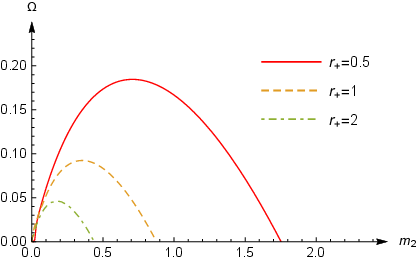}
      \hfill%
    \includegraphics{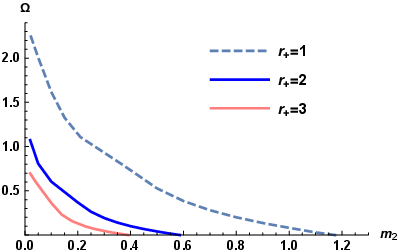}
\caption{(Left) $\Omega$ graphs as function of mass parameter $m_2$ for a linearized Ricci tensor with $r_+=0.5,~1,~2$.
The thresholds of GL instability ($\Omega=0$) are located at $m_2^{\rm th}$=1.752,~0.876,~0438. (Right) $\Omega$ graphs as function of mass parameter $m_2$ for a linearized scalar
with  $r_+=1,~2,~3$. Here, one reads off the thresholds of tachyonic  instability $m_2^{\rm sth}$ from the
points that curves of $\Omega$ intersect the positive $m_2$-axis: $m_2^{\rm sth}$=1.174, 0.587, 0.391.}
\end{figure*}

Considering  $\int^\infty_{r_+} dr [V_{\rm S}(r)/(1-r_+/r)]<0$,
one may introduce  a sufficient condition of tachyonic instability for a mass parameter $m_2$~\cite{Doneva:2017bvd}
\begin{equation}
m^2_2r_+^2<\frac{12}{10} \Rightarrow m_2<m_2^{\rm sc}=\frac{1.095}{r_+}. \label{mass-b}
\end{equation}
However, Eq.(\ref{mass-b}) is not a necessary and sufficient condition for tachyonic instability.
Observing  (Right) Fig. 1, one finds that the negative region becomes wide and deep as the mass parameter $m_2$ decreases, implying tachyonic  instability of the Schwarzschild black hole.

To determine the threshold of tachyonic instability, one has to solve the second-order differential equation (\ref{scalar-eq}) with $\omega=i\Omega$ numerically,
which may  allow an exponentially growing mode of  $e^{\Omega t} $ as  an unstable mode.
In this case,  we choose two boundary conditions: a normalizable
solution of $u(\infty)\sim e^{-\Omega r_*}$  at infinity  and
a solution of $u(r_+)\sim \left(r-r_+\right)^{\Omega r_+}$  near the horizon.
By observing  (Right) Fig.~2 together with $r_+=1$, we read off the
 bound for tachyonic instability  as
\begin{equation}
 m_2<m_2^{\rm sth}=1.174 \label{mass-c}
\end{equation}
which implies that   the threshold of tachyonic instability is given by   1.174 being greater than 1.095 (sufficient condition for tachyonic  instability).
This corresponds to a bifurcation point between Schwarzschild and $n=0$ branch of scalarized black holes. In the limit of $m^2_2 \to 0$, one has an infinitely negative potential which implies a large $\Omega$ as seen from (Right) Fig. 2.

Finally, we obtain an inequality bound for threshold of GL and tachyonic instabilities as
\begin{equation}
m_2^{\rm th}<m_2^{\rm sth}.
\end{equation}
However, we remind the reader    that the linearized  Ricci-tensor $\delta R_{\mu\nu}$ carries with a regular mass term ($m^2_2$), whereas the linearized scalar $\delta \phi$ has a tachyonic  mass term ($-1/m^2_2$).
In this sense, the GL instability is quite different from the tachyonic instability~\cite{Myung:2018iyq}.

\section{Discussions}

In this work, we have investigated  two instabilities of Schwarzschild  black holes simultaneously  by introducing   the EWS theory with a quadratic scalar coupling  to  Weyl term. Here,  the linearized  Ricci-tensor has a regular mass term ($m^2_2$), whereas the linearized scalar possesses a tachyonic  mass term ($-1/m^2_2$).
The linearized scalar equation around black hole indicates tachyonic instability for $m_2<1.174$, while  the Lichnerowicz equation for linearized  Ricci-tensor shows  GL instability for $m_2<0.876$.
This suggests that their  mass terms play  different roles  for generating scalarized black holes because the GL instability is quite different from the tachyonic instability.
We expect that the former may induce {\it infinite branches} ($n=0,1,2,\cdots$) of scalarized black holes, while the latter admits {\it single branch} ($m_2>0$) of scalarized black holes.

Now, we would like to   mention the non-Schwarzschild black hole solutions  obtained from the Einstein-Weyl theory ($\phi=0$ EWS theory with $m_2^2>0$). This solution can be obtained numerically by requiring the no-hair theorem for Ricci scalar ($R=0$)~\cite{Lu:2017kzi}.
Actually, it corresponds to single branch of non-Schwarzschild black holes with Ricci-tensor hair~\cite{Lu:2015cqa}. Recently, it was shown that  the long-wave length  instability bound for non-Schwarzschild black holes is given by $m_2<0.876$~\cite{Held:2022abx}, which is the same bound as  the GL instability for Schwarzschild black hole~\cite{Myung:2018iyq}, but it contradicts to the conjecture from black hole thermodynamics  addressed in~\cite{Lu:2017kzi,Stelle:2017bdu}.  We expect that a single branch of non-Schwarzschild black holes with Ricci-tensor and scalar hairs would be  found from the EWS theory with $f(\phi)=1+\phi^2$.

On the other hand, we consider the scalar equation (\ref{lin-eq2}) with tachyonic mass. From its static equation with $\omega=0$,   we obtain an infinite spectrum of parameter $m_2$ : $m_2\in [1.174=m_2^{\rm sth}$, 0.453, 0.280, 0.202, · · ·], which defines infinite branches of scalarized black holes: $n=0((0,1.174]),~n=1((0,0.453]),~n=2((0,0.28]),~n=3((0,0.202]),\cdots$. Also, $n=0,~1,~2,~3,\cdots$ are  identified with the number of nodes for $\delta \phi(z) = u(z)/z$ profile.
Thus, it is expected that infinite branches ($n=0,~1,~2,~3,\cdots$) of black hole with scalar hair would be found when  solving Eqs.(\ref{equa1}) and (\ref{s-equa}) numerically.
However, this computation seems not to be easy because Eq.(\ref{equa1}) includes fourth-order derivatives and its Ricci scalar is not zero ($R=2(\partial \phi)^2+\Gamma^\mu~_\mu/m^2_2$).

We wish to introduce a conventional case of $f(\phi)=\phi^2$ quadratic  coupling function. In this case, there is no GL instability because the Bach tensor-term does not contribute to the linearized Einstein equation (\ref{lin-eq1}).
Here, the linearized EWS theory reduces to the linearized EGBS theory which provides $n=0$ band with bandwidth
of $1.174 < m_2 < 1.272$ ~\cite{Silva:2017uqg}.  This band of  black holes with scalar hair is unstable against radial perturbations~\cite{Blazquez-Salcedo:2018jnn}.  This is reason why we choose  the EWS theory with the quadratic coupling function  $f(\phi)=1+\phi^2$.

Finally, for the EWS theory with a quartic coupling function $f(\phi)=(1-e^{-\kappa \phi^4})/4\kappa$~\cite{Doneva:2021tvn,Blazquez-Salcedo:2022omw,Lai:2023gwe}, the linearized scalar equation leads to $\bar{\nabla}^2\delta \phi=0$, which implies that there is no tachyonic instability. Also, its linearized Einstein equation is given by $ \delta G_{\mu\nu}=0$ which indicates that there is no GL instability. In this quartic coupling case, the linearized EWS theory reduces to the linearized EGBS theory, showing tachyonic stability.  Without tachyonic instability, one expects to  have  a single branch of nonlinearly scalarized black holes but not infinite branches of scalarized black holes.

 \vspace{1cm}

{\bf Acknowledgments}

 \vspace{1cm}
The author thanks De-Cheng Zou for helpful discussions.
 
\end{document}